\documentclass[prb,twocolumn,amsmath,amssymb,showpacs]{revtex4}

\usepackage{graphicx}
\usepackage{dcolumn}
\usepackage{bm}
\usepackage{color}

\begin{document}

\title{Large Negative Thermal Expansion in Pentacene due to Steric Hindrance}

\author{S. Haas}
\email{shaas@phys.ethz.ch}

\author{B. Batlogg}
\affiliation{Laboratory for Solid State Physics, ETH Zurich, 8093 Zurich, Switzerland
}

\author{C. Besnard}
\author{M. Schiltz}
\affiliation{Laboratory of Crystallography 1, EPF Lausanne, 1015 Lausanne, Switzerland}

\author{C. Kloc}
\author{T. Siegrist}
\email{tsi@bell-labs.com}
\affiliation{Bell Laboratories, Lucent Technologies, Murray Hill 07974, NJ, USA}

\date{\today}

\begin{abstract}

The uniaxial negative thermal expansion in pentacene crystals along $a$ is a particularity in the series of the oligoacenes, and exeptionally large for a crystalline solid. 
Full x-ray structure analysis from 120\,K to 413\,K reveals that the dominant thermal motion is  a libration of the rigid molecules about their long axes, modifying the intermolecular angle which describes the herringbone packing within the layers.
This herringbone angle increases with temperature (by 0.3\,--\,0.6$^{\circ}$ per 100\,K), and causes an anisotropic rearrangement of the molecules within the layers, i.e. an expansion in the $b$ direction, and a distinct contraction along $a$.
Additionally, a larger herringbone angle improves the cofacial overlap between adjacent, parallel molecules, and thus enhances the attractive van der Waals forces.

\end{abstract}

\maketitle

\section{\label{intro}Introduction}

The thermal expansion observed in most solids is a hallmark of anharmonic interatomic or intermolecular potentials. Whereas a hard, steep potential results in relatively small thermal expansion, a softer, shallow potential leads to substantial thermal expansion, e.g. in organic molecular materials, where the molecules are held together  by weak van der Waals (vdW) forces. Thus the thermal expansion is caused by changes of the arrangement of the molecules rather than by the small temperature-dependence of the intramolecular (covalent) bonds.

On the other hand, {\em negative\/} thermal expansion (i.e. contraction) can not be explained by a simple anharmonic two-center potential. Quite a few inorganic materials show either volume or uniaxial negative thermal expansion (NTE) \cite{Barrera2005} with ZrW$_2$O$_8$ as the most prominent example of large, isotropic NTE over a wide temperature range \cite{NTE1, NTE2, Ramirez1998}. In $\alpha$-ZrW$_2$O$_8$ (low temperature phase) and $\beta$-ZrW$_2$O$_8$ (high temperature phase), the linear expansion coefficient $\alpha$ is $-8.7\times10^{-6}$\,K$^{-1}$ and $-4.9\times10^{-6}$\,K$^{-1}$, respectively \cite{Sleight1998}.
In contrast, only very few examples of (uniaxial) NTE in organic systems are known so far (see Ref. \onlinecite{Birkedal2002} and refs. therein).

In general, NTE is caused by special geometrical arrangements of the atoms/molecules and peculiarities of the bonding which restrict the thermal movement of the atoms/molecules such that NTE occurs for geometrical reasons.
Often, such materials are characterized by open, underconstrained lattices or by a network of rigid, underconstrained entities. For example, rotation of the rigid WO$_4$ building blocks causes the overall contraction in ZrW$_2$O$_8$ \cite{Ramirez1998}.
For organic materials, different mechanisms may be involved. For instance, polyethylene \cite{White1984} contracts {\em along\/} the covalent chains as a consequence of increased vibrations perpendicular to the chains, i.e. due to motions in direction of weak vdW interactions \cite{Tasumi1965, Bruno1998}. Another mechanism has been observed for a dipeptide (TrpGly$\cdot$H$_2$O) crystal \cite{Birkedal2002}: The ordering of water molecules (located within the supramolecular peptide helices) with decreasing temperatures is responsible for the uniaxial NTE.

The intermolecular potential and its anharmonicity determine both thermal expansion and the details of thermal libration and translation motions. Both of them are accessible to experimental probes. To this end, we have performed full x-ray structure analyses of pentacene single crystals in the temperature range from 120--400\,K. 
The thermal expansion ellipsoid is calculated for pentacene and the other acene crystals, and is highly anisotropic. Pentacene crystals contract upon warming along one direction which is close to the in-plane $a$ direction. The microscopic origin of NTE is elucidated by an analysis of the refined atomic displacement parameters in terms of molecular motions. Anomalously large librations about the long molecular axis at elevated temperatures are observed, while the other librations and translations are mostly suppressed due to steric hindrance within the dense herringbone packed layers. As a consequence, the herringbone angle changes, resulting in an expansion approximately along $b$ and a contraction along $a$.

\section{\label{exp}Experimental}

Single crystals of pentacene were grown in a temperature gradient {\em in vacuo}, or by physical vapor phase transport with high purity argon as the transport gas \cite{Kloc1997, Laudise1998}. 
Pentacene crystals were investigated between room temperature and 400\,K, using an {\sc Enraf/Nonius CAD4} diffractometer (CuK$_\alpha$) with a home-made temperature control system, and additionally between 120\,K and 413\,K on the Swiss-Norwegian Beamline (SNBL, $\lambda=0.71$\,\AA) at ESRF (Grenoble).

Full structure data for pentacene single crystals at different temperatures are available from the CCDC with ref.no. xxxxxx--yyyyyy or upon request from the authors.
Crystallographic data for naphthalene, anthracene and tetracene were taken from the literature \cite{BDNaph, Oddershede2004, BDAc, RuTc2006}.
 
The thermal expansion was calculated from the change of the unit cell parameters with temperature. 
As the various acenes crystallize in different, non-orthogonal unit cells, a direct comparison of the unit cell parameters is of limited value. 
More appropriate is a transformation of the thermal expansion along the crystal axes ($\alpha_i$) into an orthogonal reference system, yielding the direction and the magnitude of the thermal expansion ellipsoid main axes $\alpha_i'$ (principal axes, eigenvalues of the thermal expansion tensor), following a procedure described earlier \cite{ohashi, finger}. For the thermal expansion analysis, the CAD4 data are considered.

\begin{figure}[tpb!]
  \centering
  \includegraphics[width=0.9\columnwidth]{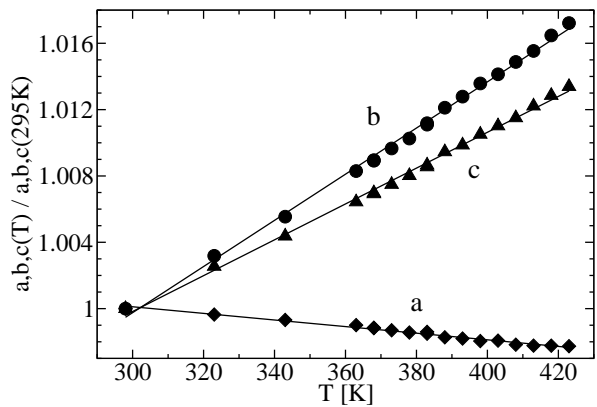}
  \caption{Temperature dependence of the lattice constants of pentacene (single crystal, CuK$_\alpha$-data). Expansion in the $ab$-plane is strongly anisotropic and {\it negative\/} along $a$. (This data corresponds to the open symbols in Fig. \ref{strain_alle}.)}
  \label{PcGitter}
\end{figure}

Additionally, molecular libration and translation parameters for pentacene were obtained by carrying out a TLS analysis \cite{Schomaker1968, TLSfootnote} from the full structure data (synchrotron data, refined with anisotropic atomic displacement parameters), using {\sc Platon} \cite{platon}. 
This analysis is carried out in the molecular inertial system and assumes rigid molecules.
B\"urgi {\it et al.\/} showed the amplitudes of the intramolecular modes to be small compared with the rigid body motion \cite{buergi}. Additionally, molecular dynamics studies indicate that in the energy range of interest the excited modes are to a high degree of intermolecular type \cite{Valle2004b}.

\section{\label{results}Results and Discussion}

\begin{figure}[tpb!]
  \centering
  \includegraphics[width=0.9\columnwidth]{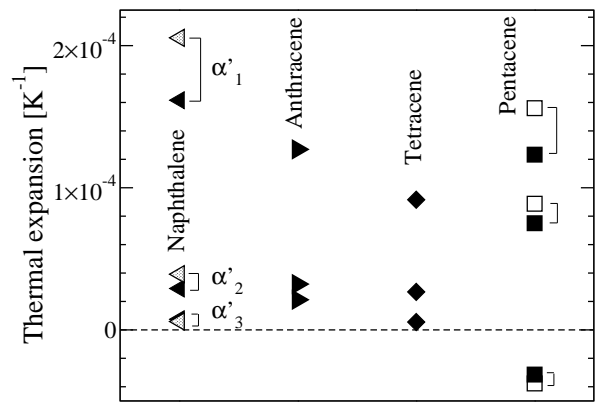}
  \caption{ The eigenvalues of the expansion tensor ($\alpha_i'$) for crystals of acenes with $n$= 2, 3, 4, and 5 benzene rings. In pentacene the expansion coefficient $\alpha_3'$ is negative. The eigenvectors are not parallel to the crystal axes (except for one direction along the $b$ axis in the monoclinic systems.) The two datasets for pentacene ($\blacksquare$, $\square$) are both measured with CuK$_\alpha$ and point detector, but with different temperature controls. Data for naphthalene taken from Oddershede {\it et al.} \cite{Oddershede2004} and Brock {\it et al.} \cite{BDNaph}(light symbols), for anthracene from Brock {\it et al.} \cite{BDAc} and for tetracene from Ref.~\onlinecite{RuTc2006}.}
  \label{strain_alle}
\end{figure}

For all investigated crystals, a structure identical with the `"H'" polymorph \cite{PcHolmes,Mattheus2001,PcSiegrist} was observed.
A plot of the unit cell axes in pentacene reveals a {\em negative\/} thermal expansion coefficient $\alpha_a$ along $a$ (Fig.~\ref{PcGitter}).
A contraction of the unit cell along $a$ can also be seen in the powder data of Mattheus {\it et al.} \cite{Mattheus2003}, although the single crystal data in Ref.~\onlinecite{Mattheus2001} would not indicate NTE.
The thermal expansion coefficients $\alpha_i'$ (expressed in orthogonal principal axes \cite{ohashi, finger}) are plotted in Fig.~\ref{strain_alle} for two pentacene samples and the shorter acenes (Refs.~\onlinecite{Oddershede2004,BDNaph,BDAc,RuTc2006}).
Most outstanding is the behavior of pentacene: the thermal expansion in one direction is distinctly negative ($-35\pm 6$$\times$$10^{-6}$\,K$^{-1}$), and positive in the two other directions.
The trend for naphthalene, anthracene and tetracene indicates a dependence of the largest thermal expansion coefficient on the length of the molecule. 
There is at least one direction with a value close to zero in all three. 
Pentacene is anomalous among the series not only in terms of the NTE, but also when its relatively large volume expansion coefficient (=$\sum \alpha_i'$) is considered.

\begin{figure}[tpb!]
  \centering
  \includegraphics[width=\columnwidth]{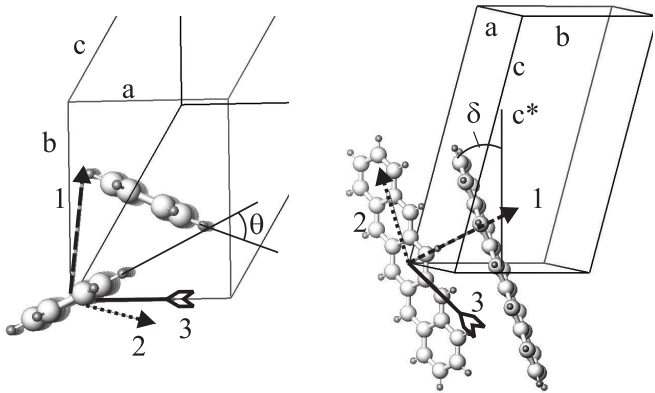}
  \caption{ Thermal expansion in pentacene: The main axes of the thermal expansion ellipsoid     are non-parallel to all crystal axes. Axis 3 (eigenvector to the value of minimal
  expansion) corresponds to a {\em negative} expansion, i.e. contraction in that direction       nearly parallel to $a$, whereas maximum expansion directs about along $b$ and towards the      center molecule (axis 1). For numerical values, see Fig. \ref{strain_alle}. $\theta$ is the    herringbone angle, and $\delta$ the tilt angle.}
  \label{eigenvectors}
\end{figure}

Fig.~\ref{eigenvectors} shows the direction of the principal axes of the thermal expansion ellipsoid with respect to the pentacene molecules and the unit cell. 
The direction of most pronounced NTE (axis 3, $\alpha_3'$) encloses an angle of 22$^{\circ}$ with the crystallographic axis $a$  and is nearly perpendicular to the crystallographic axes $b$ and $c$.
The direction of maximal expansion (axis 1, $\alpha_1'$) encloses an angle of 26$^{\circ}$ with $b$ and 50$^{\circ}$ with $c$, being almost perpendicular to $a$. 
The anomalously large intermediate eigenvalue (axis 2, $\alpha_2'$) corresponds to a direction out of the $ab$-plane, but is not exactly parallel to the long molecule axis.

To explore the microscopic cause of the negative thermal expansion in pentacene, we further analyze the temperature dependence of the thermal motions and the orientation of the molecules (herringbone/tilt angle, definitions see Fig. \ref{eigenvectors}).

\begin{table}
\caption{\label{TLS_alle} Translation and libration parameters for linear acenes at room temperature, calculated in the molecules' inertial axes coordinate system.
Libration parameters $L_{ii}$ are in deg$^2$, translation parameters $T_i$ in
$10^{-2}$\,\AA$^2$. For the monoclinic acenes, the two molecules in the unit cell
are crystallographically identical. In case of triclinic symmetry, they are
independent, creating additional parameters $L'_{ii}$ and $T'_i$ for the second
molecule. The main axes of the inertial tensor, $I_i$, are given in \AA$^2$$\cdot$amu, and also normalized with the number $n$ of benzene rings. $I_1$ (smallest momentum) points along the long molecule axis.}
\begin{ruledtabular}
\begin{tabular}{ccccccc}
($n$)&Naph\footnotemark[1] (2)&Ac\footnotemark[2] (3)&Tc\footnotemark[3] (4)&Pc (5)\\
\hline
Symmetry&mono&mono&tri&tri\\
\hline
$I_1$ ($\div n$)&130 (65)&189 (63)&252 (63)&311 (62)\\
$I_2$ ($\div n$)&356 (178)&1001 (334)&2154 (539)&3957 (791)\\
$I_3$ ($\div n$)&486 (243)&1190 (397)&2406 (602)&4268 (854)\\
\hline
$L_{11}$&19.32&15.3&11.07&9.55\\
$L_{22}$&12.18&7.2&4.51&2.69\\
$L_{33}$&16.55&10.3&3.71&2.17\\
$L'_{11}$&&&13.01&9.59\\
$L'_{22}$&&&5.25&2.76\\
$L'_{33}$&&&4.19&2.06\\
\hline
$T_1$&3.18&4.77&5.461&4.240\\
$T_2$&2.37&3.02&2.991&2.288\\
$T_3$&1.58&1.94&2.632&2.205\\
$T'_1$&&&4.636&4.504\\
$T'_2$&&&2.876&2.408\\
$T'_3$&&&2.646&2.069\\
\end{tabular}
\end{ruledtabular}
\footnotetext[1]{Values from Ref.~\cite{BDNaph}.}
\footnotetext[2]{Values from Ref.~\cite{BDAc}.}
\footnotetext[3]{Values from Ref.~\cite{RuTc2006}.}
\end{table}

The libration and translation parameters for the rigid molecule motion were calculated from synchrotron data and are listed in Table~\ref{TLS_alle}.
The longer acene molecules show the expected increase in the anisotropy of the molecules' inertial tensor. 
The longer the molecule the more suppressed (``frustrated'') are the librations due to steric hindrance, and the more closely does $L_{11}$ coincide with the molecular inertial axis 1 (given by $I_1$). In light of the negligibly small off-diagonal elements of the libration tensor (given by the coupling between the motions), $L_{11}$ describes a movement independent from $L_{22/33}$, and $L_{11}$ is aligned with $I_1$ {\em only\/} in the case of pentacene (misalignment $<$ 5$^\circ$) and tetracene ($<$7$^\circ$). The translations $T_i$, however, remain almost constant with increasing number ($n$) of benzene rings (Tab. \ref{TLS_alle}). 

\begin{figure}[tpb]
  \centering
  \includegraphics[width=0.9\columnwidth]{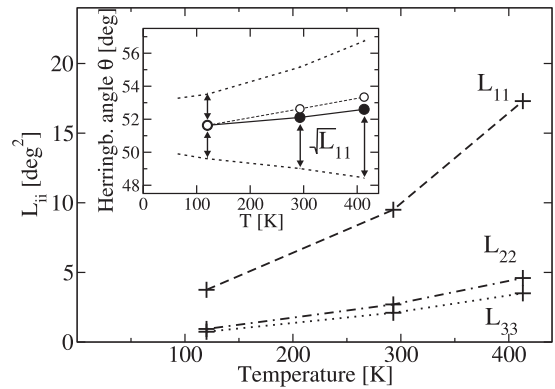}
  \caption{ Temperature dependence of the molecular librations in pentacene. The libration about the long molecule axis, $L_{11}$, is most pronounced, and strongly increases with temperature. The mean rotational amplitude is approximately $\sqrt{L_{ii}}$, i.e. $>3^\circ$ about the long axis at room temperature. Inset: The measured (static) herringbone angle $\theta$ ($\bullet$) and the mean thermal libration angle $\pm \sqrt{L_{11}}$ (connected by the broken line) as a function of temperature. 
If the thermal expansion along $b$ was caused by a {\em static\/} rotation only, the corresponding angle ($\circ$) varied more rapidly than the measured herringbone angle $\theta$ ($\bullet$).  
  }
  \label{LvsT}
\end{figure}

The herringbone angle $\theta$ (see Fig.~\ref{eigenvectors} for definition) is a central parameter to describe the packing within the layers, and is essentially the same for all acenes (51--53$^\circ$ at room temperatur), irrespective of the symmetry of the unit cell, and shows the same qualitative temperature dependence. With increasing temperature, the herringbone angle increases by 0.3--1$^\circ$ per 100\,K, with a more pronounced change in naphthalene and anthracene than in tetracene and pentacene.

\begin{figure}[tpb]
  \centering
  \includegraphics[width=0.9\columnwidth]{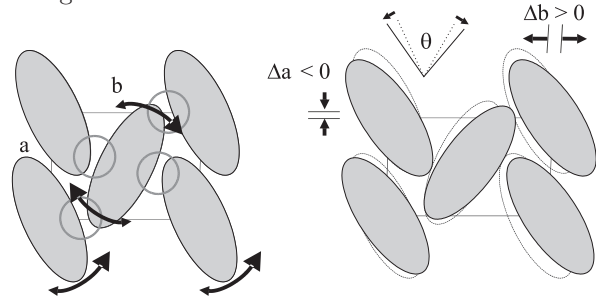}
  \caption{Schematic view of the anisotropic expansion in the $a$,$b$ plane. The libration along the long molecular axis, $L_{11}$, together with a change of the herringbone angle, are responsible for the anisotropic thermal expansion, with $a$ contracting and $b$ expanding. The other thermal libration and translation motions are strongly suppressed (``frustrated'') by steric hindrance. }
  \label{schema}
\end{figure}

The increase of the herringbone angle and the increasingly dominant librations $L_{11}$ about the long molecular axis suggest a distinctly anisotropic expansion in the $ab$-plane, as sketched in Fig.~\ref{schema}. 
The librating molecules at the corners of the unit cell interact with the one in the center, pushing each other further apart in $b$ direction as the libration amplitude grows with temperature. Thus, steric hindrance between the herringbone-packed molecules is central to the NTE in pentacene.

It is noteworthy that this simplified picture of one dominant libration $L_{11}$ is most appropriate for pentacene and tetracene, where the inertial tensor is anisotropic enough to align $L_{11}$ with the long molecular axis within 5--7$^\circ$.
Lattice dynamics calculations indicate that the lowest energy modes involve essentially pure rotational and translational {\em inter\/}molecular motions, whereas {\em intra\/}molecular excitations require higher energy (e.g. Tab. II in Ref. \onlinecite{Valle2004b}).
Thus Fig. \ref{schema} captures the molecular motions most relevant for thermal expansion in pentacene.

The observed difference of a more anisotropic (and even negative) thermal expansion in pentacene compared to tetracene can be qualitatively understood as the result of two counteracting types of thermal motion: in-plane translations and the librations along the long molecular axis. The libration apparently dominates in pentacene.

Assuming an effective width of the molecules of 7\,\AA, the thermal expansion along $b$, $\Delta b_{\rm calc}$, originating from the change of the {\em static\/} (i.e. average) herringbone angle can be easily calculated geometrically, as suggested in Fig. \ref{schema}. (The tilt of the molecules with respect to $b$ has been taken into account.) Since $\Delta b_{\rm calc} < \Delta b_{\rm exp}$, there is obviously an additional contribution from {\em dynamic\/} effects, i.e. the librations are partially out-of-phase. In a different picture, this corresponds to lattice phonons with $q \not= 0$. Consequently, an {\em effective\/} herringbone angle $\theta_{\rm eff}$ was calculated, which varies more rapidly with temperature than the measured, ``averaged'' herringbone angle $\theta$, as shown in the inset of Fig. \ref{LvsT}. Taken together, the information in Fig.~\ref{LvsT} reflects the anharmonic potential sensed by the thermal motion.

An additional factor for the negative expansion is the increased vdW force along $a$ due to an increased cofacial overlap of the pentacene molecules:
The tilt angle $\delta$ against $c^{\star}$ (definition in Fig.~\ref{eigenvectors}) decreases from $\bar{\delta}$=25.4$^{\circ}$ at 120\,K over $\bar{\delta}$=24.7$^{\circ}$ at 293\,K to $\bar{\delta}$=24.0$^{\circ}$ at 413\,K (averaged for the two inequivalent, quasi-parallel molecules). 
As a consequence, the displacement of the pentacene molecules is slightly reduced from 2.10\,\AA\ at 120\,K over 2.05\,\AA\ at 295\,K to 1.98\,\AA\ at 413\,K, and thus the cofacial overlap and the attractive vdW forces along $a$ increase.

It can be expected that the asymmetry and the uniaxial negative expansion is even more pronounced in longer herringbone packed acenes, e.g. hexacene.
In particular the effect of increasing vdW forces along the short unit cell axis, and the related change of the tilt angle are enhanced, as well as the steric hindrance within the herringbone layer.
However, the decreasing stability and the increasing ease of oxidation with increasing $n$, as well as the reported difficulties in growing hexacene crystals with reasonable quality \cite{Hexacene} make single crystal studies a challenge.

\section{Conclusions}

In summary,  the uniaxial negative thermal expansion in pentacene along $a$ is a particularity in the series of the acenes. Pentacene contracts along $a$ as a result of the increasing herringbone angle ($\Delta\theta$=0.3\,--\,0.6$^{\circ}$ per 100\,K), supported by the frustation of larger librations and translations perpendicular to the long molecule axis due to steric hindrance. Furthermore, a reduction of the relative shift of the molecules along $a$ with temperature increases the attractive vdW forces along $a$.

\section*{Acknowledgments}

We gratefully acknowledge stimulating discussions with P. Pattison, D. Chernyshov, and A. P. Ramirez.  We thank the  Swiss-Norwegian beamline (SNBL) consortium for providing access to synchrotron radiation. This work was partially funded by the U.S. Department of Energy's Nanoscale Science, Engineering, and Technology program, under grant DE-FG02-04ER46118, and the Swiss National Science Foundation.


\begin{thebibliography}{27}
\expandafter\ifx\csname natexlab\endcsname\relax\def\natexlab#1{#1}\fi
\expandafter\ifx\csname bibnamefont\endcsname\relax
  \def\bibnamefont#1{#1}\fi
\expandafter\ifx\csname bibfnamefont\endcsname\relax
  \def\bibfnamefont#1{#1}\fi
\expandafter\ifx\csname citenamefont\endcsname\relax
  \def\citenamefont#1{#1}\fi
\expandafter\ifx\csname url\endcsname\relax
  \def\url#1{\texttt{#1}}\fi
\expandafter\ifx\csname urlprefix\endcsname\relax\def\urlprefix{URL }\fi
\providecommand{\bibinfo}[2]{#2}
\providecommand{\eprint}[2][]{\url{#2}}

\bibitem[{\citenamefont{Barrera et~al.}(2005)\citenamefont{Barrera, Bruno,
  Barron, and Allan}}]{Barrera2005}
\bibinfo{author}{\bibfnamefont{G.~D.} \bibnamefont{Barrera}},
  \bibinfo{author}{\bibfnamefont{J.~A.~O.} \bibnamefont{Bruno}},
  \bibinfo{author}{\bibfnamefont{T.~H.~K.} \bibnamefont{Barron}},
  \bibnamefont{and} \bibinfo{author}{\bibfnamefont{N.~L.} \bibnamefont{Allan}},
  \bibinfo{journal}{J. Phys.: Condens. Matter} \textbf{\bibinfo{volume}{17}},
  \bibinfo{pages}{R217} (\bibinfo{year}{2005}).

\bibitem[{\citenamefont{Mary et~al.}(1996)\citenamefont{Mary, Evans, Vogt, and
  Sleight}}]{NTE1}
\bibinfo{author}{\bibfnamefont{T.~A.} \bibnamefont{Mary}},
  \bibinfo{author}{\bibfnamefont{J.~S.~O.} \bibnamefont{Evans}},
  \bibinfo{author}{\bibfnamefont{T.}~\bibnamefont{Vogt}}, \bibnamefont{and}
  \bibinfo{author}{\bibfnamefont{A.~W.} \bibnamefont{Sleight}},
  \bibinfo{journal}{Science} \textbf{\bibinfo{volume}{272}},
  \bibinfo{pages}{90} (\bibinfo{year}{1996}).

\bibitem[{\citenamefont{Evans et~al.}(1996)\citenamefont{Evans, Mary, Vogt,
  Subramanian, and Sleight}}]{NTE2}
\bibinfo{author}{\bibfnamefont{J.~S.~O.} \bibnamefont{Evans}},
  \bibinfo{author}{\bibfnamefont{T.~A.} \bibnamefont{Mary}},
  \bibinfo{author}{\bibfnamefont{T.}~\bibnamefont{Vogt}},
  \bibinfo{author}{\bibfnamefont{M.~A.} \bibnamefont{Subramanian}},
  \bibnamefont{and} \bibinfo{author}{\bibfnamefont{A.~W.}
  \bibnamefont{Sleight}}, \bibinfo{journal}{Chem. Mater.}
  \textbf{\bibinfo{volume}{8}}, \bibinfo{pages}{2809} (\bibinfo{year}{1996}).

\bibitem[{\citenamefont{Ramirez and Kowach}(1998)}]{Ramirez1998}
\bibinfo{author}{\bibfnamefont{A.~P.} \bibnamefont{Ramirez}} \bibnamefont{and}
  \bibinfo{author}{\bibfnamefont{G.~R.} \bibnamefont{Kowach}},
  \bibinfo{journal}{Phys. Rev. Lett.} \textbf{\bibinfo{volume}{80}},
  \bibinfo{pages}{4903} (\bibinfo{year}{1998}).

\bibitem[{\citenamefont{Sleight}(1998)}]{Sleight1998}
\bibinfo{author}{\bibfnamefont{A.~W.} \bibnamefont{Sleight}},
  \bibinfo{journal}{Annu. Rv. Mater. Sci.} \textbf{\bibinfo{volume}{28}},
  \bibinfo{pages}{29} (\bibinfo{year}{1998}).

\bibitem[{\citenamefont{Birkedal et~al.}(2002)\citenamefont{Birkedal,
  Schwarzenbach, and Pattison}}]{Birkedal2002}
\bibinfo{author}{\bibfnamefont{H.}~\bibnamefont{Birkedal}},
  \bibinfo{author}{\bibfnamefont{D.}~\bibnamefont{Schwarzenbach}},
  \bibnamefont{and} \bibinfo{author}{\bibfnamefont{P.}~\bibnamefont{Pattison}},
  \bibinfo{journal}{Angew. Chem. Int. Ed.} \textbf{\bibinfo{volume}{41(5)}},
  \bibinfo{pages}{754} (\bibinfo{year}{2002}).

\bibitem[{\citenamefont{White and Choy}(1984)}]{White1984}
\bibinfo{author}{\bibfnamefont{G.~K.} \bibnamefont{White}} \bibnamefont{and}
  \bibinfo{author}{\bibfnamefont{C.~L.} \bibnamefont{Choy}},
  \bibinfo{journal}{J. Polym. Sci. Polym. Phys. Edn.}
  \textbf{\bibinfo{volume}{22 (5)}}, \bibinfo{pages}{835 }
  (\bibinfo{year}{1984}).

\bibitem[{\citenamefont{Tasumi and Simanouchi}(1965)}]{Tasumi1965}
\bibinfo{author}{\bibfnamefont{M.}~\bibnamefont{Tasumi}} \bibnamefont{and}
  \bibinfo{author}{\bibfnamefont{T.}~\bibnamefont{Simanouchi}},
  \bibinfo{journal}{J. Chem. Phys.} \textbf{\bibinfo{volume}{43}},
  \bibinfo{pages}{1245} (\bibinfo{year}{1965}).

\bibitem[{\citenamefont{Bruno et~al.}(1998)\citenamefont{Bruno, Allan, Barron,
  and Turner}}]{Bruno1998}
\bibinfo{author}{\bibfnamefont{J.~A.~O.} \bibnamefont{Bruno}},
  \bibinfo{author}{\bibfnamefont{N.~L.} \bibnamefont{Allan}},
  \bibinfo{author}{\bibfnamefont{T.~H.~K.} \bibnamefont{Barron}},
  \bibnamefont{and} \bibinfo{author}{\bibfnamefont{A.~D.}
  \bibnamefont{Turner}}, \bibinfo{journal}{Phys. Rev. B}
  \textbf{\bibinfo{volume}{58 (13)}}, \bibinfo{pages}{8416 }
  (\bibinfo{year}{1998}).

\bibitem[{\citenamefont{Kloc et~al.}(1997)\citenamefont{Kloc, Simpkins,
  Siegrist, and Laudise}}]{Kloc1997}
\bibinfo{author}{\bibfnamefont{C.}~\bibnamefont{Kloc}},
  \bibinfo{author}{\bibfnamefont{P.~G.} \bibnamefont{Simpkins}},
  \bibinfo{author}{\bibfnamefont{T.}~\bibnamefont{Siegrist}}, \bibnamefont{and}
  \bibinfo{author}{\bibfnamefont{R.~A.} \bibnamefont{Laudise}},
  \bibinfo{journal}{J. Cryst. Growth} \textbf{\bibinfo{volume}{182(3-4)}},
  \bibinfo{pages}{416} (\bibinfo{year}{1997}).

\bibitem[{\citenamefont{Laudise et~al.}(1998)\citenamefont{Laudise, Kloc,
  Simpkins, and Siegrist}}]{Laudise1998}
\bibinfo{author}{\bibfnamefont{R.~A.} \bibnamefont{Laudise}},
  \bibinfo{author}{\bibfnamefont{C.}~\bibnamefont{Kloc}},
  \bibinfo{author}{\bibfnamefont{P.~G.} \bibnamefont{Simpkins}},
  \bibnamefont{and} \bibinfo{author}{\bibfnamefont{T.}~\bibnamefont{Siegrist}},
  \bibinfo{journal}{J. Cryst. Growth} \textbf{\bibinfo{volume}{187(3-4)}},
  \bibinfo{pages}{449} (\bibinfo{year}{1998}).

\bibitem[{\citenamefont{Brock and Dunitz}(1982)}]{BDNaph}
\bibinfo{author}{\bibfnamefont{C.~P.} \bibnamefont{Brock}} \bibnamefont{and}
  \bibinfo{author}{\bibfnamefont{J.~D.} \bibnamefont{Dunitz}},
  \bibinfo{journal}{Acta Crystallogr. B} \textbf{\bibinfo{volume}{38}},
  \bibinfo{pages}{2218} (\bibinfo{year}{1982}).

\bibitem[{\citenamefont{Oddershede and Larsen}(2004)}]{Oddershede2004}
\bibinfo{author}{\bibfnamefont{J.}~\bibnamefont{Oddershede}} \bibnamefont{and}
  \bibinfo{author}{\bibfnamefont{S.}~\bibnamefont{Larsen}},
  \bibinfo{journal}{J. Phys. Chem. A} \textbf{\bibinfo{volume}{108}},
  \bibinfo{pages}{1057} (\bibinfo{year}{2004}).

\bibitem[{\citenamefont{Brock and Dunitz}(1990)}]{BDAc}
\bibinfo{author}{\bibfnamefont{C.~P.} \bibnamefont{Brock}} \bibnamefont{and}
  \bibinfo{author}{\bibfnamefont{J.~D.} \bibnamefont{Dunitz}},
  \bibinfo{journal}{Acta Crystallogr. B} \textbf{\bibinfo{volume}{46}},
  \bibinfo{pages}{795} (\bibinfo{year}{1990}).

\bibitem[{\citenamefont{Haas et~al.}()\citenamefont{Haas, Kloc, and
  Siegrist}}]{RuTc2006}
\bibinfo{author}{\bibfnamefont{S.}~\bibnamefont{Haas}},
  \bibinfo{author}{\bibfnamefont{C.}~\bibnamefont{Kloc}}, \bibnamefont{and}
  \bibinfo{author}{\bibfnamefont{T.}~\bibnamefont{Siegrist}}, \bibinfo{note}{to
  be published}.

\bibitem[{\citenamefont{Ohashi and Burnham}(1973)}]{ohashi}
\bibinfo{author}{\bibfnamefont{Y.}~\bibnamefont{Ohashi}} \bibnamefont{and}
  \bibinfo{author}{\bibfnamefont{C.~W.} \bibnamefont{Burnham}},
  \bibinfo{journal}{Am. Mineral.} \textbf{\bibinfo{volume}{58}},
  \bibinfo{pages}{843} (\bibinfo{year}{1973}).

\bibitem[{\citenamefont{Hazen and Finger}(1982)}]{finger}
\bibinfo{author}{\bibfnamefont{R.~M.} \bibnamefont{Hazen}} \bibnamefont{and}
  \bibinfo{author}{\bibfnamefont{L.~W.} \bibnamefont{Finger}},
  \emph{\bibinfo{title}{Comparative crystal chemistry}}
  (\bibinfo{publisher}{John Wiley}, \bibinfo{year}{1982}).

\bibitem[{\citenamefont{Schomaker and Trueblood}(1968)}]{Schomaker1968}
\bibinfo{author}{\bibfnamefont{V.}~\bibnamefont{Schomaker}} \bibnamefont{and}
  \bibinfo{author}{\bibfnamefont{K.~N.} \bibnamefont{Trueblood}},
  \bibinfo{journal}{Acta Crystallogr. B} \textbf{\bibinfo{volume}{24}},
  \bibinfo{pages}{63} (\bibinfo{year}{1968}).

\bibitem[{TLS()}]{TLSfootnote}
\bibinfo{note}{The molecular motion is described in the molecular inertial
  system by a translation ($\mathbf T$) and a libration ($\mathbf L$) tensor,
  together with a coupling tensor $\mathbf S$.}

\bibitem[{\citenamefont{Spek}(2001)}]{platon}
\bibinfo{author}{\bibfnamefont{A.~L.} \bibnamefont{Spek}},
  \emph{\bibinfo{title}{Platon, a multipurpose crystallographic tool}},
  \bibinfo{howpublished}{Utrecht University, Utrecht} (\bibinfo{year}{2001}),
  \bibinfo{note}{http://www.cryst.chem.uu.nl/platon/}.

\bibitem[{\citenamefont{B\"urgi et~al.}(2001)\citenamefont{B\"urgi,
  Rangavittal, and Hauser}}]{buergi}
\bibinfo{author}{\bibfnamefont{H.-B.} \bibnamefont{B\"urgi}},
  \bibinfo{author}{\bibfnamefont{N.}~\bibnamefont{Rangavittal}},
  \bibnamefont{and} \bibinfo{author}{\bibfnamefont{J.}~\bibnamefont{Hauser}},
  \bibinfo{journal}{Helv. Chim. Acta} \textbf{\bibinfo{volume}{84}},
  \bibinfo{pages}{1889} (\bibinfo{year}{2001}).

\bibitem[{\citenamefont{Della-Valle et~al.}(2004)\citenamefont{Della-Valle,
  Venuti, Farina, Brillante, Masino, and Girlando}}]{Valle2004b}
\bibinfo{author}{\bibfnamefont{R.}~\bibnamefont{Della-Valle}},
  \bibinfo{author}{\bibfnamefont{E.}~\bibnamefont{Venuti}},
  \bibinfo{author}{\bibfnamefont{L.}~\bibnamefont{Farina}},
  \bibinfo{author}{\bibfnamefont{A.}~\bibnamefont{Brillante}},
  \bibinfo{author}{\bibfnamefont{M.}~\bibnamefont{Masino}}, \bibnamefont{and}
  \bibinfo{author}{\bibfnamefont{A.}~\bibnamefont{Girlando}},
  \bibinfo{journal}{J. Phys. Chem. B} \textbf{\bibinfo{volume}{108}},
  \bibinfo{pages}{1822} (\bibinfo{year}{2004}).

\bibitem[{\citenamefont{Holmes et~al.}(1999)\citenamefont{Holmes, Kumaraswamy,
  Matzger, and Vollhardt}}]{PcHolmes}
\bibinfo{author}{\bibfnamefont{D.}~\bibnamefont{Holmes}},
  \bibinfo{author}{\bibfnamefont{S.}~\bibnamefont{Kumaraswamy}},
  \bibinfo{author}{\bibfnamefont{A.~J.} \bibnamefont{Matzger}},
  \bibnamefont{and} \bibinfo{author}{\bibfnamefont{K.~P.~C.}
  \bibnamefont{Vollhardt}}, \bibinfo{journal}{Chem.-Eur. J.}
  \textbf{\bibinfo{volume}{5}}, \bibinfo{pages}{3399} (\bibinfo{year}{1999}).

\bibitem[{\citenamefont{Mattheus et~al.}(2001)\citenamefont{Mattheus, Dros,
  Baas, Meetsma, de~Boer, and Palstra}}]{Mattheus2001}
\bibinfo{author}{\bibfnamefont{C.~C.} \bibnamefont{Mattheus}},
  \bibinfo{author}{\bibfnamefont{A.~B.} \bibnamefont{Dros}},
  \bibinfo{author}{\bibfnamefont{J.}~\bibnamefont{Baas}},
  \bibinfo{author}{\bibfnamefont{A.}~\bibnamefont{Meetsma}},
  \bibinfo{author}{\bibfnamefont{J.~L.} \bibnamefont{de~Boer}},
  \bibnamefont{and} \bibinfo{author}{\bibfnamefont{T.~T.~M.}
  \bibnamefont{Palstra}}, \bibinfo{journal}{Acta Crystallogr. C}
  \textbf{\bibinfo{volume}{57}}, \bibinfo{pages}{939} (\bibinfo{year}{2001}).

\bibitem[{\citenamefont{Siegrist et~al.}(2001)\citenamefont{Siegrist, Kloc,
  Sch\"on, Batlogg, Haddon, Berg, and Thomas}}]{PcSiegrist}
\bibinfo{author}{\bibfnamefont{T.}~\bibnamefont{Siegrist}},
  \bibinfo{author}{\bibfnamefont{C.}~\bibnamefont{Kloc}},
  \bibinfo{author}{\bibfnamefont{J.~H.} \bibnamefont{Sch\"on}},
  \bibinfo{author}{\bibfnamefont{B.}~\bibnamefont{Batlogg}},
  \bibinfo{author}{\bibfnamefont{R.~C.} \bibnamefont{Haddon}},
  \bibinfo{author}{\bibfnamefont{S.}~\bibnamefont{Berg}}, \bibnamefont{and}
  \bibinfo{author}{\bibfnamefont{G.~A.} \bibnamefont{Thomas}},
  \bibinfo{journal}{Angew. Chem. Int. Ed.} \textbf{\bibinfo{volume}{40}},
  \bibinfo{pages}{1732} (\bibinfo{year}{2001}).

\bibitem[{\citenamefont{Mattheus et~al.}(2003)\citenamefont{Mattheus, Dros,
  Baas, Oostergetel, Meetsma, de~Boer, and Palstra}}]{Mattheus2003}
\bibinfo{author}{\bibfnamefont{C.~C.} \bibnamefont{Mattheus}},
  \bibinfo{author}{\bibfnamefont{A.~B.} \bibnamefont{Dros}},
  \bibinfo{author}{\bibfnamefont{J.}~\bibnamefont{Baas}},
  \bibinfo{author}{\bibfnamefont{G.~T.} \bibnamefont{Oostergetel}},
  \bibinfo{author}{\bibfnamefont{A.}~\bibnamefont{Meetsma}},
  \bibinfo{author}{\bibfnamefont{J.~L.} \bibnamefont{de~Boer}},
  \bibnamefont{and} \bibinfo{author}{\bibfnamefont{T.~T.~M.}
  \bibnamefont{Palstra}}, \bibinfo{journal}{Synthetic Metals}
  \textbf{\bibinfo{volume}{138}}, \bibinfo{pages}{475} (\bibinfo{year}{2003}).

\bibitem[{\citenamefont{Campbell et~al.}(1962)\citenamefont{Campbell,
  Robertson, and Trotter}}]{Hexacene}
\bibinfo{author}{\bibfnamefont{R.~B.} \bibnamefont{Campbell}},
  \bibinfo{author}{\bibfnamefont{J.~M.} \bibnamefont{Robertson}},
  \bibnamefont{and} \bibinfo{author}{\bibfnamefont{J.}~\bibnamefont{Trotter}},
  \bibinfo{journal}{Acta Crystallogr.} \textbf{\bibinfo{volume}{15}},
  \bibinfo{pages}{289} (\bibinfo{year}{1962}).

\end{thebibliography}
\end{document}